\documentstyle[prl,floats,aps,twocolumn,epsf,graphicx]{revtex}
\begin{document}
\twocolumn[\hsize\textwidth\columnwidth\hsize\csname
@twocolumnfalse\endcsname

\title{High-Order Corrections to the Entropy and Area of Quantum Black Holes}
\author{Shahar Hod}
\address{The Racah Institute of Physics, The Hebrew University, Jerusalem 91904, Israel}
\date{\today}
\maketitle

\begin{abstract}

\ \ \ The celebrated area-entropy formula for black holes has provided 
the most important clue in the search for the elusive theory of quantum gravity. 
We explore the possibility that the (linear) area-entropy relation acquires some smaller 
corrections.  Using the Boltzmann-Einstein formula, we rule out the 
possibility for a power-law correction, and provide severe constraints 
on the coefficient of a possible log-area correction. We argue that a non-zero 
logarithmic correction to the area-entropy relation, would also 
imply a modification of the area-mass relation for quantum black holes.

\end{abstract}
\bigskip

]

The necessity in a quantum theory of gravity was already 
recognized in the 1930s. However, despite the flurry of research 
we still lack a complete theory of quantum gravity.
It is believed that black holes may play a major role in our 
attempts to shed some light
on the nature of a quantum theory of gravity (such as the role 
played by atoms in the early development of quantum mechanics). 

In particular, the area-entropy relation $S_{BH}=A/4\ell_P^2$ \cite{Beken1} 
for black holes has served as a valuable element of guidance for the 
quantum-gravity research. The intuition that has led Bekenstein to this 
discovery is actually based on very simple ingredients. In particular, 
to elucidate the relation between area and entropy, it is instructive to
use a semiclassical version of Christodoulou's reversible 
processes \cite{Chris,ChrisRuff}, in which a particle is absorbed by a black hole. 
Bekenstein \cite{Beken1,Beken2} has shown that the {\it Heisenberg quantum uncertainty principle} 
imposes a lower bound on the increase in black hole surface area

\begin{equation}\label{Eq1}
(\Delta A)_{\min} = \gamma \ell_P^2\  ,
\end{equation}
where $\gamma$ is a dimensionless constant of order unity, and 
$\ell_P=\left({G \over {c^3}}\right)^{1/2} {\hbar}^{1/2}$ is the
Planck length. 
Remarkably, this bound is {\it universal} in the sense that it is {\it independent} of the 
black-hole parameters \cite{Note1}. 
The universality of the fundamental lower bound 
is clearly a strong evidence in favor of 
a {\it uniformly} spaced area spectrum for quantum
black holes (see Ref. \cite{Beken3}). Hence, one concludes that 
the quantization condition of the black-hole surface area 
should be of the form

\begin{equation}\label{Eq2}
A_n=\gamma {\ell^2_P} \cdot n\ \ \ ;\ \ \ n=1,2,\ldots\  ,
\end{equation}

Furthermore, using the fact that the minimum increase in black-hole 
surface area should correspond to a minimum increase of its entropy (in 
order to compensate for the loss of the particle's entropy), one arrives to 
the {\it proportionality} between black-hole surface area and entropy $S_{BH}=\eta A/\ell_P^2$.

It should be recognized however that the precise values of the proportionality 
constants $\gamma$ and $\eta$ cannot be inferred from 
this simple line of reasoning. The very nature of Heisenberg quantum uncertainty principle allows only 
an order-of-magnitude estimate of the minimal increase in black-hole 
surface area. As a consequence, the proportionality constant $\eta$ was fixed only few 
years later by Hawking, who determined the characteristic temperature of black holes \cite{Haw}.
 
Mukhanov and Bekenstein \cite{Muk,BekMuk,Beken3} have suggested an independent 
argument in order to determine the value of the coefficient $\gamma$. 
In the spirit of Boltzmann-Einstein formula in
statistical physics, they relate $g_n \equiv exp[S_{BH}(n)]$ to the number of microstates of the
black hole that correspond to a particular external macrostate. 
In other words, $g_n$ is the degeneracy of the $n$th area eigenvalue. 
The thermodynamic relation between black-hole surface area and entropy can
be met with the requirement that $g_n$ has to be an integer for every
$n$ only when
 
\begin{equation}\label{Eq3}
\gamma =4\ln{k} \  ,
\end{equation}
where $k$ is some natural number. Thus, statistical physics arguments force 
the dimensionless constant $\gamma$ in Eq. (\ref{Eq2}) to 
be of the form Eq. (\ref{Eq3}). 

Nevertheless, a specific value of $k$ requires further input. This information 
may emerge by applying {\it Bohr's correspondence principle} \cite{Hod2} to 
the (discrete) quasinormal frequencies of black holes \cite{Nollert,Andersson}. 
This argument provides the missing link, and gives evidence in favor of the value $k=3$. It should be
mentioned that following the pioneering work of Bekenstein
\cite{Beken1}, a number of independent calculations (most of them in
the last few years) have indicated that a black-hole surface area has a {\it discrete} 
spectrum. Moreover, many of them have recovered the uniformly spaced area
spectrum Eq. (\ref{Eq2}) \cite{Kogan,Maggiore,Lousto,Peleg,LoukoMakela,BarKun,Makela,Kastrup}.
However, there is no general agreement on the spacing of the levels. 
The relation $\gamma =4 \ln 3$ is 
the {\it unique} value consistent both with the area-entropy
{\it thermodynamic} relation, with {\it statistical physics} arguments (namely, 
the Boltzmann-Einstein formula), and with {\it Bohr's correspondence principle.}

Moreover, we would like to emphasize that using Bohr's correspondence principle allows one to 
fix not only the value of $k$, but also to obtain the factor of $1 \over 4$ in the linear 
area-entropy relation (a factor which could not be fixed from Bekenstein's 
intuitive argument \cite{Beken1}, and which was derived only later from Hawking's analysis 
of black-hole radiation \cite{Haw}).

While it is well established that the leading term in the area-entropy relation is linear 
in the black-hole surface area, in recent years evidence has been mounting that 
smaller correction terms may also exist (see e.g. \cite{Amel} and references therein). 
These indications for sub-leading {\it logarithmic} terms have arised both in String Theory and 
in Loop Quantum Gravity. 
One should emphasize that there is, however, {\it no} general 
agreement on the {\it coefficient} of the logarithmic correction \cite{Amel}. 
It is therefore highly important to establish constraints on the possible values that
these sub-leading entropy corrections may take. To that end, 
we consider a general area-entropy relation for black holes of the form 

\begin{equation}\label{Eq4}
S_{bh}={1 \over 4} {A \over \ell_P^2} + \alpha_1 \Big({A \over \ell_P^2}\Big)^{\beta} + \ln [f(A)]\  ,
\end{equation}
where $\beta > 0$, and $f(A)$ is a function of the black-hole surface area. 
In addition, we write the quantized black-hole surface area in the form

\begin{equation}\label{Eq5}
A_n=(\gamma_0 n + \gamma_1 n^{\delta} +\gamma_2 \ln{n}) \ell_P^2 \  ,
\end{equation}
where $\delta>0$ and $n=1,2,\ldots$ \cite{Note2}. 
For $g_n$ to be an integer for every $n$, $f(A)$ should equal a natural number for 
every $A_n$, implying that $f(A)$ should be of the form $f(A)=\sum_{j=0}^{j_{max}} c_j A^j$, such 
that all powers in the series are natural numbers. In the large area limit we expand 
$\ln[f(A)]=\alpha_2 \ln A +\alpha_3 +\cdots$. 
Substituting Eq. (\ref{Eq5}) in Eq. (\ref{Eq4}), and using the 
Boltzmann-Einstein relation $g_n=\exp[S_{bh}(n)]$ (with the requirement that $g_n$ is an 
integer for every $n$), one obtains severe constrains on the possible values that 
the various coefficients may take. First, the Boltzmann-Einstein formula implies that 
$\gamma_0$ should be of the form \cite{Muk,BekMuk,Beken3,Hod2}

\begin{equation}\label{Eq6}
\gamma_0=4\ln{k}\  ,
\end{equation}
where $k$ is a natural number. In addition, we find

\begin{equation}\label{Eq7}
\alpha_1=\gamma_1=0\  .
\end{equation}
Thus, our simple argument implies that there are {\it no} stronger-than-logarithmic corrections to the 
area-entropy relation. To continue, one has to consider two distinct cases:

{\it Case I. Non-vanishing logarithmic corrections to the area-entropy relation} $(\alpha_2 \neq 0)$. 
In this case one finds that the coefficients must satisfy the constraints 

\begin{equation}\label{Eq8}
\gamma_2=0\  ,
\end{equation}
and 

\begin{equation}\label{Eq9}
\alpha_2=l\ \ \ ; \ \ \ \alpha_3=\ln\Big({m \over {{\gamma_0}^l}}\Big)\  ,
\end{equation}
where $l$ and $m$ are natural numbers. Thus, the black-hole surface area 
takes a very simple form

\begin{equation}\label{Eq10}
A_n=4\ell_P^2\ln{k} \cdot n\ ;\ \ \ n=1,2,\ldots\  .
\end{equation}
Such a {\it uniformly} spaced area spectrum (with {\it no} sub-leading corrections) 
supports the existence of a {\it fundamental} area unit \cite{Note3}. In addition, the 
area-entropy relation should be of the restricted form

\begin{equation}\label{Eq11}
S_{bh}=S_{BH} +l\cdot\ln\Big({A \over \ell_P^2}\Big)+\ln\Big[{m \over {(4\ln{k})^l}}\Big]\  .
\end{equation}
The main conclusion is that the coefficient of the log-area correction should be 
a {\it natural} number.

{\it Case II. No logarithmic corrections to the area-entropy relation} ($\alpha_2=0$). 
In this case one finds that the coefficients must satisfy the constraints 

\begin{equation}\label{Eq12}
\gamma_2=4l\ \ \ ; \ \ \ \alpha_3=\ln{m}\  ,
\end{equation}
where $l$ and $m$ are natural numbers.

In summary, using a simple argument based on the Boltzmann-Einstein formula, we have 
derived severe constraints on the possible sub-leading corrections to the (semiclassical) 
Bekenstein-Hawking area-entropy relation for black holes. In particular, we have ruled out the possibility for a 
stronger-than-logarithmic correction, and found that the coefficient of a possible logarithmic correction should be 
a natural number. 

\bigskip
\noindent
{\bf ACKNOWLEDGMENTS}
\bigskip

This research was supported by G.I.F. Foundation.


\begin{thebibliography}{99}

\bibitem{Beken1} J. D. Bekenstein, Phys. Rev. D {\bf 7}, 2333 (1973).

\bibitem{Chris} D. Christodoulou, Phys. Rev. Lett. {\bf 25}, 1596 (1970).

\bibitem{ChrisRuff} D. Christodoulou and R. Ruffini, Phys. Rev. D {\bf 4}, 3552 (1971).

\bibitem{Beken2} J. D. Bekenstein, Lett. Nuovo Cimento {\bf 11}, 467 (1974).

\bibitem{Note1} The universal lower bound Eq. (\ref{Eq1}) derived by Bekenstein is
valid only for {\it neutral} particles \cite{Beken1,Beken2}. The assimilation of a quantum 
{\it charged} particle by a black hole yields a similar lower bound for 
the increase in black-hole surface area $(\Delta A)_{\min}=4 {\ell_P^2}$ \cite{Hod1}. 
As was noted by Bekenstein \cite{Beken1,Beken2} (for neutral particles), the underling physics 
which excludes a completely reversible process is the {\it Heisenberg quantum uncertainty principle}. 
However, for {\it charged} particles it must be supplemented by another physical 
mechanism \cite{Hod1} -- a Schwinger discharge of the black hole 
({\it vacuum polarization} effects). Without this physical mechanism, one 
could have reached the reversible limit ($\Delta A=0$). 
It is remarkable that the lower bound found for charged particles is 
of the same order of magnitude as the one given by
Bekenstein for neutral particles, even though they 
emerge from {\it different} physical mechanisms.
This is clearly a strong evidence in favor of a {\it uniformly} spaced area spectrum 
for quantum black holes.

\bibitem{Hod1} S. Hod, Phys. Rev. D {\bf 59} 024014 (1999).

\bibitem{Beken3} J. D. Bekenstein in XVII Brazilian National Meeting
  on Particles and Fields, eds. A. J. da Silva et. al. (Brazilian
  Physical Society, Sao Paulo, 1996), J. D. Bekenstein in Proceedings of the VIII
  Marcel Grossmann Meeting on General Relativity, eds. T. Piran and
  R. Ruffini (World Scientific , Singapore, 1998).

\bibitem{Haw} S. W. Hawking, Commun. Math. Phys. {\bf 43}, 199 (1975).

\bibitem{Muk} V. Mukhanov, JETP Lett. {\bf 44}, 63 (1986).

\bibitem{BekMuk} J. D. Bekenstein and V. F. Mukhanov, Phys. Lett. B {\bf 360}, 7 (1995).

\bibitem{Hod2} S. Hod, Phys. Rev. Lett. {\bf 81} 4293 (1998).

\bibitem{Nollert} H-P. Nollert, Phys. Rev. D {\bf 47}, 5253 (1993).

\bibitem{Andersson} N. Andersson, Class. Quantum Grav. {\bf 10}, L61 (1993).

\bibitem{Kogan} Ya. I. Kogan, JETP Lett. {\bf 44}, 267 (1986).

\bibitem{Maggiore} M. Maggiore, Nucl. Phys. B {\bf 429}, 205 (1994).

\bibitem{Lousto} C. O. Lousto, Phys. Rev. D {\bf 51}, 1733 (1995).

\bibitem{Peleg} Y. Peleg, Phys. Lett. B {\bf 356}, 462 (1995).

\bibitem{LoukoMakela} J. Louko and J. M\"akel\"a, Phys. Rev. D {\bf 54}, 4982 (1996).

\bibitem{BarKun} A. Barvinsky and G. Kunstatter, Phys. Lett. B {\bf 389}, 231 (1996); 
A. Barvinsky, S. Das, and G. Kunstatter, Class. Quant. Grav. {\bf 18}, 4845 (2001); 
S. Das, P. Ramadevi, U. A. Yajnik, and A .Sule, Phys. Lett. B {\bf 565}, 201 (2003); 
A. Alekseev, A. Polychronakos and M. Smedback, Phys. Lett. B {\bf 574}, 296 (2003); 
A. Polychronakos, Phys. Rev. D {\bf 69}, 044010 (2004).

\bibitem{Makela} J. M\"akel\"a, preprint gr-qc/9602008.

\bibitem{Kastrup} H. Kastrup, Phys. Lett. B {\bf 385}, 75 (1996).

\bibitem{Amel} G. Amelino-Camelia, M. Arzano and A. Procaccini, e-print gr-qc/0405084; 
D. V. Fursaev, Phys. Rev. D {\bf 51}, 5352 (1995); H. Kastrup, Phys. Lett. B {\bf 413}, 267 (1997); 
M. I. Park, e-print hep-th/0402173; 
S. Das, P. Majumdar, and R. Bhaduri, Class. Quant. Grav. {\bf 19}, 2355 (2002); 
C. O. Lousto, Phys. Lett. B {\bf 352}, 228 (1995).

\bibitem{Note2} It is possible to add a constant term in Eq. (\ref{Eq5}) as well. 
It would then be straightforward to use the Boltzmann-Einstein formula in order to 
derive similar constraints on the possible values that this constant term may take.

\bibitem{Note3} Analyzing carefully the Gedanken experiment of Bekenstein \cite{Beken1}, 
or the wave analysis of \cite{Hod2}, one concludes that 
$\Delta A \sim \ell_P^2 +O({{\ell_P^4} \over {M^2}})$. The sub-leading term is a 
consequence of two distinct factors: (i) the second term in the area-mass relation 
$\Delta A=32\pi M\Delta M+16\pi(\Delta M)^2$ with $\Delta M \sim {{\ell_P^2} \over M}$, 
and (ii) the gravitational back-reaction caused by 
the particle's energy $E \sim {{\ell_P^2} \over M}$ 
(in Bekenstein's analysis), or the gravitational wave  energy in \cite{Hod2}, which 
would change the effective black-hole mass in these analyses from $M$ to 
$M + O({{\ell_P^2} \over M})$. The relation 
$\Delta A \sim \ell_P^2 +O({{\ell_P^4} \over {M^2}})$ suggests that $A_n$ acquires a 
sub-leading correction term $O(\ln{n})$. We have learned, however, that {\it if} a logarithmic 
correction exist in the area-entropy relation, then the area spectrum should be uniformly 
spaced. This may indicate that the area-mass relation for quantum black holes should 
also acquire higher-order corrections of the form 
$A=16\pi M^2+\xi \ell_P^2 \ln({{M^2} \over {\ell_P^2}})$, 
where $\xi$ is a dimensionless constant of order unity. Such a relation may allow 
$\Delta A \sim \ell_P^2$ without $O({{\ell_P^4} \over {M^2}})$ corrections. 

\end{thebibliography}
\end{document}